\documentclass{article} \usepackage{slashed,feynmf,epsfig,amsmath,amssymb,enumitem} \usepackage[papersize={8.5in,11in}]{geometry}
  \renewcommand{\vec}[1]{\boldsymbol{\mathrm{#1}}} \usepackage{bm}
\usepackage{graphicx}
\begin{document}
\title{The Weak Lensing Mass of Cosmic Web Filaments and Modified Gravity (MOG)}
\author{J. W. Moffat\\
Perimeter Institute for Theoretical Physics, Waterloo, Ontario N2L 2Y5, Canada\\
and\\
Department of Physics and Astronomy, University of Waterloo, Waterloo,\\
Ontario N2L 3G1, Canada}
\maketitle



\begin{abstract}
The weak lensing of cosmic web filaments is investigated in modified gravity (MOG) and it is demonstrated that the detected galaxies and baryonic gas of order $5-15 \%$ in filaments can with the enhanced value of the strength of gravity agree with the lensing data for filaments without dark matter.
\end{abstract}

\maketitle


\section{Introduction}

The pioneering research by Zwicky~\cite{Zwicky} and Vera Rubin and collaborators~\cite{Rubin1,Rubin2} showed that the dynamics of galaxies and galaxy clusters were in disagreement with Newtonian and Einstein gravity. The masses of galaxies and clusters inferred from dynamics were found to exceed the baryon mass in these systems. This led to the postulate that dark matter forms halos around galaxies and clusters, resolving the discrepancy with gravity. There is no convincing evidence either in deep underground laboratory experiments such as LUX~\cite{LUX}, Panda-X~\cite{PandaX} and astrophysical and LHC experiments to support the existence of exotic dark matter particles. Modified gravitation is an alternative explanation for the dynamics of galaxies and clusters. A fully covariant and relativistic theory of gravitation called Scalar-Tensor-Vector-Gravity (STVG) or MOG (modified gravity)
~\cite{Moffat1} fits galaxy rotation curves and galaxy dynamics~\cite{MoffatRahvar1,MoffatToth1,Haghighi1,Roshan1,Roshan2,Brownstein}, describes successfully the dynamics of clusters~\cite{MoffatRahvar2,Haghighi2,MoffatToth2} and merging clusters, such as the Bullet Cluster and the Train Wreck Cluster Abell 520~\cite{BrownsteinMoffat,IsraelMoffat} without detectable dark matter in the present Universe. The theory can also fit lensing data~\cite{MoffatRahvarToth} without dark matter.

MOG also describes successfully early Universe cosmology. The density of matter at the CMB, before the formation of stars and galaxies, is dominated in MOG by the density of massive spin 1, gravitationally sourced, electrically neutral vector bosons $\phi_\mu$~\cite{Moffat1,Moffat2,MoffatToth3,Shojai}, which explains the growth of structure and the CMB data. At the formation of stars, galaxies and clusters the baryon density $\rho_b$ dominates, $\rho_b >>\rho_\phi$, and $G=G_N(1+\alpha)$ with $\alpha > 1$. Gravitational waves in MOG and their consequences for the advanced LIGO/Vergo detection of gravitational waves has been investigated~\cite{Moffat3}.

Wide-field galaxy redshift surveys have shown that the large-scale matter distribution in the late-time Universe is not homogeneous~\cite{Bond,Yess}. Matter falls together under the influence of gravity into filamentary structures forming the cosmic web~\cite{Springel}. A prediction of the cold dark matter model (CDM) is that a network of dark matter cosmic filaments connect dark matter halos.  In the following, we will show that the masses and densities of the filaments of the cosmic web can be explained by modified gravity (MOG) without detectable dark matter in the late-time universe.  In a recent paper~\cite{Hudson}, the weak lensing signal of an ensemble of filaments is stacked between groups and clusters of galaxies. The weak lensing signal is detected, employing CFHTLenS galaxy ellipticities, from stacked filaments between SDSS-III/BOSS luminous red galaxies (LRGs).  An excess filament mass density is detected in the projected LRG pairs at the $5\sigma$ level with a mass $(1.6\pm 0.3)\times 10^{13}M_\odot$ for a filament region $7.1\, h^{-1}\,{\rm Mpc}$ long and $2.5\, h^{-1}\,{\rm Mpc}$ wide. The claim is made that this establishes that the cosmic web filaments are composed of dark matter.  

In a separate paper~\cite{Eckert}, it has been reported that X-ray observations of filamentary structure of gas at $10^7$ kelvin associated with the galaxy cluster Abell 2744 reveal hot gas in the filamentary structure extending to a scale of 8 Mpc. It is demonstrated that cosmic web filaments coincide with over-densities of galaxies and dark matter, with $5-15\%$ of their mass in baryonic gas, and it is claimed that this accounts for a large fraction of the missing baryons in the Universe. 

We shall show in the following that MOG can explain the over-densities and masses of filaments with an enhanced gravitational strength coupling $G=G_N(1+\alpha)$, where $G_N$ denotes Newton's gravitational constant, with a non-zero parameter $\alpha$ and with a $5-15\%$ baryonic gas and no dark matter. The baryons in the filament gas can account for the missing baryons in the Universe.

\section{MOG Poisson equation and acceleration law}

In the weak gravitational field limit, the MOG field equations lead to the point particle potential for a static spherically symmetric system:
\begin{equation}
\Phi(r)=\Phi_N(r)+\Phi_Y(r),
\end{equation}
where
\begin{equation}
\Phi_N(r)=-\frac{G_{\infty}M}{r}=-\frac{G_N(1+\alpha)M}{r},
\end{equation}
and
\begin{equation}
\Phi_Y(r)=\frac{\alpha G_NM\exp(-\mu r)}{r}.
\end{equation}
For continuous distributions of matter with baryon density $\rho_{\rm bar}$, the $\Phi_N$ and $\Phi_Y$ potentials are given by
\begin{equation}
\Phi_N({\vec x})=-G_N(1+\alpha)\int d^3{\vec x}'\frac{\rho_{\rm bar}({\vec x}')}{|{\vec x}-{\vec x}'|},
\end{equation}
and
\begin{equation}
\Phi_Y({\vec x})=\alpha G_N\int d^3{\vec x}'\frac{\exp(-\mu|{\vec x}-{\vec x}'|)\rho_{\rm bar}({\vec x}')}{|{\vec x}-{\vec x}'|}.
\end{equation}
The complete MOG potential for a given density $\rho_{\rm bar}({\vec x})$ is
\begin{equation}
\Phi(\vec{x}) = - G_N \int d^3x'\frac{\rho_{\rm bar}(\vec{x}')}{|\vec{x}-\vec{x}'|}\Big[1+\alpha -\alpha \exp(-\mu|\vec{x}-\vec{x}'|)\Big].
\label{potential}
\end{equation}
For a delta-function source density $\rho_{\rm bar}({\vec x})=M\delta^3({\vec x})$.

The Poisson equations for $\Phi_N(r)$ and $\Phi_Y(r)$ are given by
\begin{equation}
\label{NewtonPot}
\nabla^2\Phi_N(r)=4\pi G\rho_{\rm bar}(r),
\end{equation}
and
\label{YukawaPot}
\begin{equation}
(\nabla^2-\mu^2)\Phi_Y(r)=-4\pi\alpha G_N\rho_{\rm bar}(r),
\end{equation}
respectively.

The modified acceleration law for point particles is given by~\cite{Moffat1}:
\begin{equation}
a_{\rm MOG}(r)=-\frac{G_NM}{r^2}\biggl[1+\alpha-\alpha\exp(-\mu r)(1+\mu r)\biggr].
\end{equation}
The acceleration law can be extended to a distribution of matter:
\begin{equation}
\label{accelerationlaw2}
a_{\rm MOG}({\vec x})=-G_N\int d^3{\vec x}'\frac{\rho_{\rm bar}({\vec x}')({\vec x}-{\vec x}')}{|{\vec x}-{\vec x}'|^3}
[1+\alpha-\alpha\exp(-\mu|{\vec x}-{\vec x}'|(1+\mu|{\vec x}-{\vec x}'|)],
\end{equation}
where $\rho_{\rm bary}(\bf x)$ is the total baryon density of matter.

\section{Weak lensing of filaments}

We will obtain a convergence $\kappa$-map by measuring the distortion of images of background sources caused by the deflection of light as it passes the cosmic web filaments (lenses). A measurement is made of the reduced shear $g=\gamma/(1-\kappa)$, where $\gamma$ is the anisotropic stretching of the filament image, and the convergence $\kappa$ is the shape-independent change in the size of the image. 

We can derive a $\kappa$-map in MOG~\cite{BrownsteinMoffat}:
\begin{equation}
\label{kappamap}
\kappa(x,y)=\int dz\frac{4\pi G(r)}{c^2}\frac{D_lD_{ls}}{D_s}\rho(x,y,z)=\frac{\Sigma(x,y)}{\Sigma_c(r)},
\end{equation}
where
\begin{equation}
\Sigma(x,y)=\int\rho(x,y,z)dz,
\end{equation}
and $\Sigma(x,y)$ is the filament surface mass density, and 
\begin{equation}
\label{kappamap2}
\Sigma_c(r)=\frac{G_N}{G(r)}\Sigma_c=\frac{\Sigma_c}{{\cal G}(r)}.
\end{equation}
is the Newtonian critical surface mass density with vanishing shear. Moreover, $D_s$ is the angular distance to a light source, $D_l$ is the angular distance to the filament lens, and $D_{ls}$ is the angular distance from the filament to the light source. The stacked LRG pairs ensemble average for $\bar\Sigma_c$ is found to be $\bar\Sigma_c=1640M_\odot/pc^2$. 

We have written the MOG weak field acceleration law as
\begin{equation}
a(r)=-\frac{G(r)M}{r^2},
\end{equation}
where
\begin{equation}
G(r)=G_N\biggl[1+\alpha-\alpha\exp(-\mu r)(1+\mu r)\biggr],
\end{equation}
and ${\cal G}(r)=G(r)/G_N$. For the length size the filament $\sim 7$ Mpc and the width size $\sim 3$ Mpc, we can take it that $G\sim G_N(1+\alpha)$ where $\alpha$ is approximately constant. The Eqs. (\ref{kappamap}) and (\ref{kappamap2}) are only valid in the thin lens approximation and should in general be replaced by
\begin{equation}
\kappa(x,y)=\int dz\frac{4\pi G(r)}{c^2}\frac{D_lD_{ls}}{D_s}\rho(x,y,z)=\frac{\tilde\Sigma(x,y)}{\Sigma_c(r)},
\end{equation}
where
\begin{equation}
\tilde\Sigma(x,y)=\int dz{\cal G}(r)\rho(x,y,z).
\end{equation}

We can now determine the filaments total mass from the scaled mass obtained from MOG:
\begin{equation}
M_T=M_b+M_{CDM}=(1+\alpha)M_b,
\end{equation}
where $M_{\rm CDM}$ is the CDM mass and $M_T=(1.6\pm 0.3)\times 10^{13}M_\odot$ is the total estimated filament mass~\cite{Hudson}.  The baryon mass is $M_b=fM_T=(2.4\pm 0.05)\times 10^{12}M_\odot$ where we have adopted the filament baryon mass fraction $f=15\%$~\cite{Eckert}. Solving for $\alpha$ we obtain:
\begin{equation}
\alpha=\frac{M_T}{M_b}-1=5.7,
\end{equation}
yielding $G=6.7G_N$.

Eckert et al.~\cite{Eckert} estimate the baryon gas fraction $f$ in filaments between $5$ and $15\%$ for the various substructures, representing a significant fraction of the universal baryon fraction of $15\%$~\cite{Planck}. The census of all observed baryons in the Universe falls short of the estimate of $5\%$ of the Universe's total energy content~\cite{Fukugita,Cen}. The baryon plasma temperature is in the range $(10-20)\times 10^6$ K for the filaments, which is significantly less than the virial temperature of the cluster Abell 2744 core considered by Eckert et al.~\cite{Eckert}. This gas temperature corresponds to those expected for the hottest parts of the warm intergalactic medium~\cite{Cen,Dave,Gheller}. Numerical simulations predict that the intergalactic filament gas should have temperatures in the range $10^{5.5} - 10^{6.5}$ Kelvin, which is significantly less than than the virial temperature of the cluster core of order $10^8$ Kelvin, making the filaments appear to be composed of dark matter.

\section{Conclusions}

In the standard model, the cosmic web filaments that connect dark matter halos are composed of dark matter. We have demonstrated that in MOG galaxies and clusters are not composed of dominant dark matter halos and their dynamics can be well described by baryons with an enhanced value of the gravitational strength $G=G_N(1+\alpha)$ without dark matter. Moreover, we have shown that the cosmic filaments can be demonstrated to contain only a baryon gas and still agree with weak lensing observations without exotic dark matter.

\section*{Acknowledgments}

I thank Martin Green for helpful discusions. This research was supported in part by Perimeter Institute for Theoretical Physics. Research at Perimeter Institute is supported by the Government of Canada through the Department of Innovation, Science and Economic Development Canada and by the and by the Province of Ontario through the Ministry of Research, Innovation and Science.

\end{document}